\documentclass[%
 reprint,
superscriptaddress,
 amsmath,amssymb,
 aps,
 pra,
]{revtex4-2}
\usepackage{graphicx}
\usepackage{dcolumn}
\usepackage{bm}
\usepackage{lipsum}
\usepackage{braket}
\usepackage{siunitx}
\usepackage{epigraph}
\usepackage{todonotes}

\begin{document}

\title{Low-energy spectrum of double-junction superconducting circuits in the Born-Oppenheimer approximation}

\author{Leo Uhre Jakobsen}
\affiliation{Center for Quantum Devices, Niels Bohr Institute, University of Copenhagen, Denmark}
\affiliation{NNF Quantum Computing Programme, Niels Bohr Institute, University of Copenhagen, Denmark}
\author{Ksenia Shagalov}
\affiliation{Center for Quantum Devices, Niels Bohr Institute, University of Copenhagen, Denmark}
\affiliation{NNF Quantum Computing Programme, Niels Bohr Institute, University of Copenhagen, Denmark}
\author{David Feldstein-Bofill}
\affiliation{Center for Quantum Devices, Niels Bohr Institute, University of Copenhagen, Denmark}
\affiliation{NNF Quantum Computing Programme, Niels Bohr Institute, University of Copenhagen, Denmark}
\author{Morten Kjaergaard}
\affiliation{Center for Quantum Devices, Niels Bohr Institute, University of Copenhagen, Denmark}
\affiliation{NNF Quantum Computing Programme, Niels Bohr Institute, University of Copenhagen, Denmark}
\author{Karsten Flensberg}
\affiliation{Center for Quantum Devices, Niels Bohr Institute, University of Copenhagen, Denmark}
\author{Svend Krøjer}
\affiliation{Center for Quantum Devices, Niels Bohr Institute, University of Copenhagen, Denmark}
\affiliation{NNF Quantum Computing Programme, Niels Bohr Institute, University of Copenhagen, Denmark}

\date{\today}

\begin{abstract}

The superconductor–insulator–superconductor Josephson junction is the fundamental nonlinear element of superconducting circuits. Connecting two junctions in series gives rise to higher-harmonic content in the total energy-phase relation, enabling new design opportunities in multimode circuits. However, the double-junction element hosts an internal mode whose spectrum is set by the finite capacitances of the individual junctions. Using a Born–Oppenheimer approximation that treats the additional mode as fast compared to the qubit mode, we analyze the double-junction circuit element shunted by a large capacitor. 
Here, we derive an effective single-mode model of the qubit containing a correction term owing to the presence of the internal mode. In experimentally relevant parameter regimes, we numerically find that our model accurately describes the low-energy spectrum of the qubit. We further discuss how eliminating the internal degree of freedom affects the system's periodic boundary conditions and leads to non-uniqueness in performing the Born-Oppenheimer approximation. 
Finally, we analyze the harmonic content of the double-junction element and discuss its sensitivity to charge noise.
\end{abstract}

\maketitle

\section{Introduction}

Superconducting circuits constitute a versatile and flexible platform for quantum metrology, quantum-limited amplification, and quantum processors~\cite{kjaergaard2020current_state_of_play, krantz2019quantum_engineers_guide, clerk2010review_measurement, blais2021cqed_review, degen2017review_sensing,esposito2021review_twpa}. The Josephson junction, typically realized as a superconductor-insulator-superconductor (SIS) junction, is the fundamental nonlinear element enabling these technologies~\cite{kim2025josephson_review, siddiqi2021review_scq}. For example, the nonlinearity of a single junction shunted by a large capacitor or a large inductor leads to the weakly anharmonic transmon qubit~\cite{koch2007transmon} and strongly anharmonic fluxonium qubit~\cite{manucharyan2009fluxonium}, respectively. Beyond these conceptually simple single-mode qubits, superconducting circuits based on multiple Josephson junctions allow for the engineering of multi-mode devices, often introducing original design opportunities. Examples count various noise-protected superconducting qubits~\cite{gyenis2021moving, doucot2012review_rhombus, bell2014rhombus_chains,brooks2013_0_pi_qubit, groszkowski2018_0_pi_coherence, dipaolo2019_0_pi_enhancement, gyenis2021_0_pi_qubit, kalashnikov2020bifluxon, smith2020two_cooper_pair_tunneling, smith2022magnifying_fluctuations_kite, hays2025harmonium, nguyen2025grid_state_qubit, feldstein_bofill2026cooper_pair_parity, roverch2026experimentalrealizationcos2varphitransmon}, readout techniques beyond conventional dispersive readout~\cite{didier2015longitudinal_readout, pfeiffer2024pmon_decoupling_qubit,chapple2025junction_readout, wang2025junction_readout,beaulieu2026junction_readout, hazra2025readout_double_junction, shagalov2025double_junction_higher_harmonics}, flexible mixing and coupling elements~\cite{frattini2017snails_3_wave_mixing, sivak2019snail_kerr_free_3_wave_mixing, abdo2013josephson_ring_modulator, christensen2023schemeparitycontrolledmultiqubitgates}, and general Josephson harmonics engineering~\cite{bozkurt2023fourier_engineering}. Here, Josephson harmonics engineering represents a particularly flexible way to use multiple junctions to, in principle, realize any energy-phase relation of the resultant two-terminal multi-mode circuit. This idea builds on the experimentally verified~\cite{banszerus2024higher_harmonics, banszerus2025rhombus_higher_harmonics, shagalov2025double_junction_higher_harmonics, feldstein_bofill2026cooper_pair_parity} insight that two junctions in series give rise to a non-sinusoidal energy-phase relation which can be used as a resource of higher Josephson harmonics.

A common obstacle when working with multi-mode circuits is that they often require approximations to reduce the number of degrees of freedom at the level of the Hamiltonian~\cite{dipaolo2019_0_pi_enhancement, smith2020two_cooper_pair_tunneling, smith2022magnifying_fluctuations_kite, hays2025harmonium, frattini2017snails_3_wave_mixing, roverch2026experimentalrealizationcos2varphitransmon}. Such simplifications are important since they increase intuition, guide execution, and interpretation of experiments and significantly reduce computational overhead in simulations. Multi-mode devices are typically designed and realized such that the low-energy physics is described by a single primary mode, while internal high-energy modes only participate indirectly. In this case, the internal modes can be ``integrated out'' to yield an effective Hamiltonian where the primary mode(s) is(are) renormalized by the additional modes~\cite{Ciani_DiVincenzo_Terhal_2025lecture_notes}. In a classical approximation, the internal modes are assumed to be fixed via energy minimization, neglecting their quantum zero-point fluctuations. This approach typically works well for harmonic modes without external capacitors, since the intrinsic capacitance of inductors is vanishingly small, leading to highly localized zero-point fluctuations and a very high excitation energy~\cite{smith2020two_cooper_pair_tunneling, smith2022magnifying_fluctuations_kite, roverch2026experimentalrealizationcos2varphitransmon}. However, for periodic modes whose inductive energy is set by Josephson junctions, the intrinsic junction capacitance can be appreciable in experiments, resulting in increased zero-point fluctuations and decreased excitation energy of the internal modes~\cite{shagalov2025double_junction_higher_harmonics, feldstein_bofill2026cooper_pair_parity}. In this case, the classical approximation is not guaranteed to be accurate, and methods that take into account the quantum dynamics of the auxiliary modes are required.

Beyond classical approaches, the Born-Oppenheimer (BO) approximation has proved useful in the study of multi-mode qubits to systematically account for zero-point fluctuations of high-energy modes~\cite{dipaolo2019_0_pi_enhancement, rymarz2023singular_quantization, hays2025harmonium,shagalov2025double_junction_higher_harmonics, feldstein_bofill2026cooper_pair_parity}. The BO approximation assumes high-energy ``fast'' modes are in the ground state such that the corresponding energy depends parametrically on the low-energy ``slow'' modes, effectively renormalizing the potential energy of the slow modes~\cite{Ciani_DiVincenzo_Terhal_2025lecture_notes, rymarz2023singular_quantization}. When the BO approximation is applied to degrees of freedom with periodic boundary conditions, such as high-energy transmon modes, it raises practical questions about how the system's periodicity should be interpreted as the number of degrees of freedom is reduced.

In this work, we consider a two-terminal double-junction circuit element and derive an effective single-mode model by integrating out its internal mode as shown schematically in Fig.~\ref{fig:fig1_circuit_intro}(a,c). We obtain the circuit's effective energy-phase relation renormalized by the internal mode via the BO approximation and perform a detailed numerical investigation of the parameter space of validity by comparing to the full two-mode model. We discuss how the periodic boundary conditions are treated in the BO approximation and find that the two-mode model can be significantly simplified before applying the BO approximation. We also analyze how the Josephson harmonics are renormalized and quantify how the charge noise sensitivity of the auxiliary mode is inherited by the primary qubit mode. With this analysis, the double-junction element can be considered as a single higher-harmonic component that can be readily utilized when developing superconducting circuits.
 
The paper is organized as follows: In Sec.~\ref{sec:background}, we introduce the double-junction circuit and review the classical single-mode model, including its harmonics. In Sec.~\ref{sec:theory}, we give a full two-mode description of the circuit. We review the BO approximation and obtain the analytical BO correction to the classical model. In Sec~\ref{sec:results}, we investigate the accuracy of the BO single-mode model, its higher Josephson harmonics, and the charge dispersion due to offset charges on the central superconducting island. In Sec.~\ref{sec:conclusion}, we summarize the results of the paper.

\section{Background}\label{sec:background}

\begin{figure}
    \centering
    \includegraphics[width=1\columnwidth]{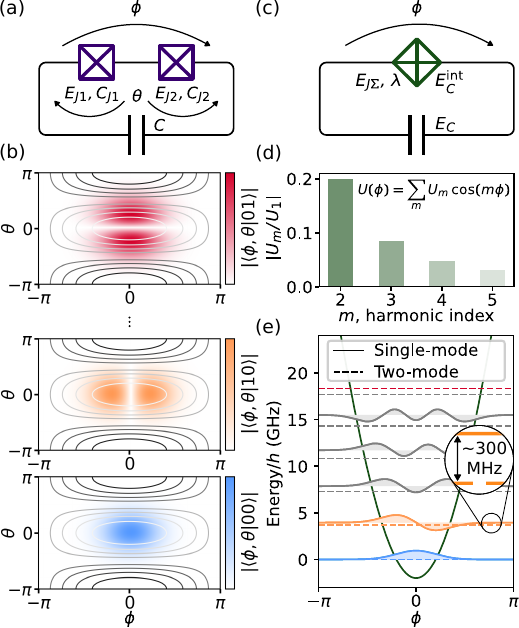}
    \caption{The double-junction circuit as a single effective element. (a) Circuit diagram of the double-junction qubit with two Josephson junctions in series, shunted by a large capacitor. A low-energy qubit mode, $\phi$, and a high-energy internal mode, $\theta$, are shown with relevant circuit parameters. (b) Phase basis wavefunctions of the ground state (blue, bottom), the first qubit excitation (orange, middle) and the first internal mode excitation (red, top). Contour lines indicate the double-junction potential (Eq.~\eqref{eq:H_qubit-internal}). (c) An effective single-element representation of the double-junction circuit describing the qubit mode using the total Josephson energy $E_{J\Sigma}$ and junction $\lambda$-parameter. (d) Appreciable higher harmonic coefficients extracted from the classical single-mode model (Eq.~\eqref{eq:U_classical}). (e) Qubit wavefunctions of the classical single-mode model (solid lines, effective potential energy in dark green) compared to the full two-mode model (dashed lines, first internal mode excitation in red), showing significant discrepancies of the classical model.
    The parameters are representative of realistic device parameters; $E_{J1}/h=E_{J2}/h=20\,\si{\giga\hertz}$, $E_C^\mathrm{int}/h=1.25\,\si{\giga\hertz}$ ($C_{J1} = C_{J2} \approx 8\,\si{\femto\farad}$), $E_C/h=200\,\si{\mega\hertz}$ ($C \approx 93\,\si{\femto\farad}$).}
    \label{fig:fig1_circuit_intro}
\end{figure}

The double-junction circuit consists of a large shunting capacitance $C$ in parallel with two junctions $E_{J1}, E_{J2}$ in series, see Fig.~\ref{fig:fig1_circuit_intro}(a). The phase difference across the two junctions constitute the qubit degree of freedom $\phi$ while the presence of the middle island gives rise to an auxiliary degree of freedom, which we refer to as the internal mode $\theta$. The frequencies of the two modes are determined by their respective charging energies, which in turn are determined by the large shunting capacitor for the qubit mode and by the smaller junction capacitances $C_{J1}, C_{J2}$ for the internal mode. The intrinsic junction capacitances are set by the junction overlap area ($100\times100\,\si{\nano\meter^2}$ scale) corresponding to junction capacitances approximately one order of magnitude less than the typical shunting capacitances used in transmon qubits~\cite{kim2025josephson_review, shagalov2025double_junction_higher_harmonics, feldstein_bofill2025gatemon_revisited}.
Hence, for experimentally relevant parameters, the charging energy of the internal mode is significantly larger than for the qubit mode. Double-junction transmons therefore have several excitations of the qubit mode at energies below the first excited state of the internal mode, see Fig.~\ref{fig:fig1_circuit_intro}(b,e). 
Consequently, an effective single-mode model that accurately describes the low-energy excitations of the qubit can be obtained by assuming the internal mode is in its ground state~\cite{Ciani_DiVincenzo_Terhal_2025lecture_notes}. 

When excitations in the internal mode are excluded, tunneling across the two junctions is correlated, and as a consequence, the two junctions in series can be considered a single effective element, see Fig.~\ref{fig:fig1_circuit_intro}(c). The correlated tunneling across this effective element admits higher-order tunneling of Cooper pairs (i.e. pairs of Cooper pairs), corresponding to higher Josephson harmonics in the single-mode energy-phase relation
\begin{equation}\label{eq:Fourier_series}
    U(\phi)=\sum_m U_m \cos(m\phi),
\end{equation}
where $U_m$ are the Josephson harmonic coefficients, see Fig.~\ref{fig:fig1_circuit_intro}(d). 
The effective energy-phase relation $U(\phi)$ can be estimated in a classical approximation where the internal mode is assumed to be fixed to the value that minimizes the classical energy~\cite{bozkurt2023fourier_engineering}, giving the result
\begin{equation}\label{eq:U_classical}
    U_\mathrm{classical}(\phi)=-E_{J\Sigma}\sqrt{1-\lambda\sin^2\frac{\phi}{2}}.
\end{equation}
Here, $E_{J\Sigma}=E_{J1}+E_{J2}$ is the sum of the Josephson energies and $\lambda=4E_{J1}E_{J2}/(E_{J1}+E_{J2})^2$ is a junction asymmetry parameter which we refer to as the junction $\lambda$-parameter. The junction $\lambda$-parameter is unity for $E_{J1}=E_{J2}$. In this \textit{balanced} case, the second harmonic of $U_\mathrm{classical}$ can become as large as $20\%$ of the fundamental harmonic, see Fig.~\ref{fig:fig1_circuit_intro}(d).

As pointed out in Ref.~\cite{bozkurt2023fourier_engineering}, there is a striking analogy between Eq.~\eqref{eq:U_classical} and the energy-phase relation of a semiconductor-based junction~\cite{beenakker1991universal}
\begin{equation}
    U_\mathrm{semi}(\phi)=-\Delta\sum_i \sqrt{1-T_i\sin^2\frac{\phi}{2}},
\end{equation}
where the junction $\lambda$-parameter plays the role of the transmission probability $T_i$ of Andreev processes and $E_{J\Sigma}$ plays the role of the superconducting gap $\Delta$. Although this analogy is attractive, it is not an exact equivalence due to the dynamics of the internal mode in the double-junction circuit, which are neglected in Eq.~\eqref{eq:U_classical}. 
By comparing the low-energy spectrum of the single-mode model in Eq.~\eqref{eq:U_classical} to the full two-mode model (Eq.~\eqref{eq:potential_phi1_phi2} below), we find a discrepancy in the qubit frequency of around $\sim300\,\si{\mega\hertz}$ for realistic parameters that put the qubit mode in the transmon regime, see Fig.~\ref{fig:fig1_circuit_intro}(e). In the next section, we start from the full two-mode model of the qubit and internal modes and perform a BO approximation to obtain a single-mode model that accounts for the renormalization due to the internal mode.

\section{Theory}\label{sec:theory}

\subsection{Two-mode model}

We start with the full two-mode Hamiltonian of the double-junction circuit as obtained from standard circuit quantization~\cite{Ciani_DiVincenzo_Terhal_2025lecture_notes, rasmussen2021review_circuit_companion}. This enables us to take into account the zero-point fluctuations of the internal mode. The two-mode Hamiltonian is
\begin{align}\nonumber
    H =&\ 4E_{C1} ( n_1-n_{g1})^2 + 4E_{C2} ( n_2-n_{g2})^2 - g  n_1  n_2 \\ \label{eq:potential_phi1_phi2}
    &- E_{J1} \cos( \varphi_1) - E_{J2} \cos( \varphi_2).
\end{align}
Here, the Hamiltonian is given in terms of the individual phase drops across the two junctions $\varphi_1, \varphi_2$ and their conjugate Cooper pair number operators $n_1, n_2$. For each Cooper pair number operator, there is an associated offset charge $n_{g1}, n_{g2}$. The charging energies are given by $E_{C1} = e^2 (C + C_{J2})/2C_\Sigma^2 $, $E_{C2} = e^2 (C + C_{J1})/2C_\Sigma^2 $, and $  g = 4e^2 C/C_\Sigma^2$ where $C_\Sigma^2=C (C_{J1}+C_{J2})+ C_{J1} C_{J2}$.

In order to obtain a single-mode description of the qubit degree of freedom, we transform the variables to diagonalize the capacitive term in the Hamiltonian. The transformation is
\begin{align} \label{eq:transformation}
\boldsymbol{\phi}'&=\mathbf{S}^T\boldsymbol{\phi}, \quad \mathbf{n}'=\mathbf{S}^{-1}\mathbf{n},\\
    \begin{pmatrix}
        \phi\\
        \theta
    \end{pmatrix}
    &=
    \begin{pmatrix}
        1 & 1\\
        \frac{k+1}{2} & \frac{k-1}{2}
    \end{pmatrix}
    \begin{pmatrix}
        \varphi_1\\
        \varphi_2
    \end{pmatrix},\\  
        \begin{pmatrix}
        n\\
        N
    \end{pmatrix}
    &=
    \begin{pmatrix}
        \frac{1-k}{2} & \frac{1+k}{2}\\
        1 & -1
    \end{pmatrix}
    \begin{pmatrix}
        n_1\\
        n_2
    \end{pmatrix},
\end{align}
where $k=(C_{J1}-C_{J2})/(C_{J1}+C_{J2})$ is the junction capacitance asymmetry. The intrinsic junction capacitances are geometrically linked to the Josephson energies, however, here $C_{J1}, C_{J2}$ collects all capacitive contributions across each junction, thus $k$ is an independent parameter. The non-orthogonal, but canonical, transformation in Eq.~\eqref{eq:transformation} is inspired by Ref.~\cite{rymarz2023singular_quantization} and preserves the qubit mode as the sum of the junction phase differences $\phi=\varphi_1+\varphi_2$. This is important as we seek a description of the two junctions as a single effective element that can be used with external circuitry. Additionally, the transformation preserves the phase space volume of phase and charge separately since $|\det \mathbf{S}|=1$.

The Hamiltonian resulting from the transformation in Eq.~\eqref{eq:transformation} is
\begin{align}\label{eq:H_qubit-internal}
    H' &= 4E_C ( n-n_g)^2 + 4E_C^\mathrm{int} ( N-N_g)^2 \\ \nonumber
    &- E_{J\Sigma} \cos\frac{\phi}{2}\cos\left(\theta-k\frac{\phi}{2}\right) + E_{J\Delta} \sin\frac{\phi}{2}\sin\left(\theta-k\frac{\phi}{2}\right),
\end{align}
where the charging energies associated with the qubit and internal modes are $E_C = e^2/2(C+C_{J1}C_{J2}/(C_{J1}+C_{J2})) $ and $E_C^\mathrm{int} = e^2/2(C_{J1}+C_{J2}) $ while $E_{J\Sigma}=E_{J1}+E_{J2}$ and $E_{J\Delta}=E_{J1}-E_{J2}$ are the sum and difference of the Josephson energies, respectively. The offset charges for the qubit and internal modes, $n_g$ and $N_g$, are suitable redefinitions of $n_{g1}, n_{g2}$. The resulting potential energy in Eq.~\eqref{eq:H_qubit-internal} is shown in Fig.~\ref{fig:fig2_2d_potentials}(a).
Despite the apparent asymmetry in the definition of $\phi$ and $\theta$ in Eq.~\eqref{eq:transformation}, the two modes can be faithfully interpreted as representing the dynamics of the qubit island and middle island; the qubit mode experiences the shunting capacitance $C$ in parallel with the series combination of the junction capacitances $C_{J1}$ and $C_{J2}$, whereas the internal mode experiences the junction capacitances in parallel.
Finally, the transformation also couples the boundary conditions of the $\phi$- and $\theta$-modes, as we discuss in Sec.~\ref{sec:simple_boundary}.

\subsection{Born-Oppenheimer approximation}

The goal is now to integrate out the internal $\theta$-mode from the transformed Hamiltonian in Eq.~\eqref{eq:H_qubit-internal} using the BO approximation. This will result in a correction term arising from the internal mode zero-point fluctuations that adds to the classically obtained energy-phase relation $U_\mathrm{classical}(\phi)$. As a reminder, the BO approximation separates the fast (internal) and slow (qubit) modes using an ansatz wavefunction and Hamiltonian of the form~\cite{dipaolo2019_0_pi_enhancement, rymarz2023singular_quantization, Ciani_DiVincenzo_Terhal_2025lecture_notes}
\begin{align}
\Psi_\mathrm{ansatz}&=\chi^\phi(\theta)\psi(\phi),\\
H_\mathrm{ansatz}&=H_\mathrm{slow}(\phi)+H_\mathrm{fast}^\phi(\theta).
\end{align}
The slow mode enters parametrically (indicated by superscript) in the ansatz wavefunction and the Hamiltonian describing the fast mode. A low-energy Hamiltonian that only describes the slow mode is then obtained by assuming that the fast mode is in its ground state
\begin{align}
H_\mathrm{ansatz}\Psi_\mathrm{ansatz}&=\left[H_\mathrm{slow}(\phi)+H_\mathrm{fast}^\phi(\theta)\right]\chi^\phi(\theta)\psi(\phi)\\
    \longrightarrow H_\mathrm{BO}\,\psi(\phi)&= \left[H_\mathrm{slow}(\phi)+U_\mathrm{corr}(\phi)\right]\psi(\phi),
\end{align}
where the BO correction potential $U_\mathrm{corr}$ is the ground state energy of $H_\mathrm{fast}$ with $\phi$ promoted to a quantum variable again. 

In the next section, we discuss the implications for the coupled, periodic boundary conditions as the BO approximation separates the slow and fast modes.

\subsection{Decoupling modes and boundary conditions}\label{sec:simple_boundary}

\begin{figure}
    \centering
    \includegraphics[width=1\columnwidth]{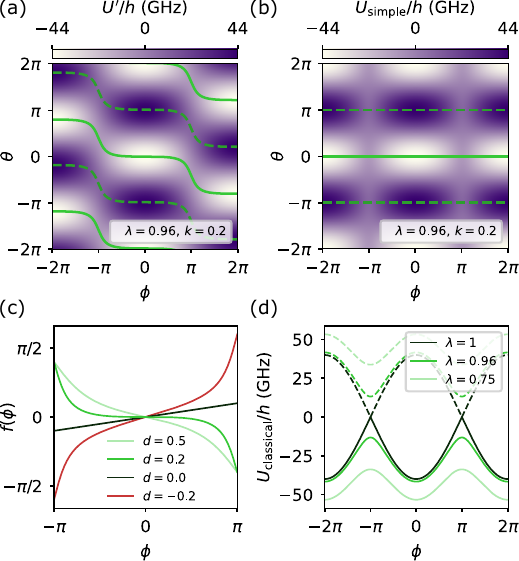}
    \caption{Potential energy of (a) the double-junction circuit (Eq.~\eqref{eq:H_qubit-internal}) and (b) the simplified potential (Eq.~\eqref{eq:simple_potential}) for junction asymmetry $\lambda = 0.96$ ($d=0.2$). The lines of minimum and maximum potential energy as a function of $\phi$ are shown in solid and dashed green, respectively. The simplified potential preserves the potential in the $\theta$-direction while the line of minima is fixed to $\theta=0$. (c) Plot of the function $f(\phi)$ (Eq.~\eqref{eq:f_phi}) for fixed capacitance asymmetry $k=0.2$ and varying junction asymmetry $d$.
    (d) The potential energy along the minimum (solid) and maximum (dashed) lines from (a-b) for different values of the junction $\lambda$-parameter. We fix $k=0.2$, $E_C/h=200$\,\si{\mega\hertz}, $\lambda E_{J\Sigma}/4 = 50E_C$ and $E_{J\Sigma}/E_C^\mathrm{int}=32$ as $\lambda$ is varied.
    }
    \label{fig:fig2_2d_potentials}
\end{figure}

The original boundary conditions are $2\pi$-periodic in $\varphi_1$ and $\varphi_2$ which translates to a coupled boundary condition for the qubit and internal modes
\begin{equation}\label{eq:coupled_boundary_condition}
    (\phi, \theta)=\left(\phi+2\pi,\theta\pm\pi+\pi k\right),
\end{equation}
for $\varphi_{1}\to\varphi_{1}+2\pi$ (positive sign) or $\varphi_{2}\to\varphi_{2}+2\pi$ (negative sign). As the BO approximation separates the slow and fast modes, their coupled boundary conditions also decouple, and the resulting potential $U_\mathrm{BO}(\phi)=U_\mathrm{classical}(\phi)+U_\mathrm{corr}(\phi)$ preserves the original $2\pi$-periodicity of the $\phi$-mode independent of the boundary condition of the $\theta$-mode. This implies that the single-mode BO potential $U_\mathrm{BO}$ can arise from a family of two-mode ansatz potentials with different coupled boundary conditions. To see this explicitly, we rewrite the potential in Eq.~\eqref{eq:H_qubit-internal} using standard trigonometric identities
\begin{align}\label{eq:U_prime}
    U'(\phi,\theta)&=- s(\phi) E_{J\Sigma}\cos\left(\theta-f(\phi)\right)\sqrt{1-\lambda\sin^2\frac{\phi}{2}},\\ \label{eq:f_phi}
    f(\phi)&=k\frac{\phi}{2}-\arctan\left(d\tan\frac{\phi}{2}\right),
\end{align}
where $s(\phi)=\mathrm{sign}[\cos(\phi/2)]$ is a sign and $d=E_{J\Delta}/E_{J\Sigma}$ is the junction asymmetry related to the junction $\lambda$-parameter $\lambda=\sqrt{1-d^2}$. See Fig.~\ref{fig:fig2_2d_potentials}(a) for a plot of $U'$ for $k=d=0.2$ ($\lambda=0.96$).

In the BO approximation, when $\phi$ is reduced to a parameter and the fast mode is assumed to be in its ground state, both parameters $f^\phi$ and $s^\phi$ are absorbed by the fast mode, which always localizes at its minimum $\theta_\mathrm{min}^\phi=f^\phi+\pi(1-s^\phi)/2$ (solid green lines in Fig.~\ref{fig:fig2_2d_potentials}(a,b), dashed green lines correspond to the maximum). This means that the BO approximation erases information about the minimum of the fast mode, and any (well-behaved) function $f(\phi)$ will therefore result in the same single-mode BO potential $U_\mathrm{BO}(\phi)$ despite representing different two-mode potentials. We can use this fact to understand two aspects of the BO approximation; first, to decouple the boundary conditions of the two modes by simplifying the potential and, second, to disentangle errors introduced by the BO approximation.

We start by choosing a simpler potential $U_\mathrm{simple}$ in place of $U'$ and use it to compute the BO correction in Sec.~\ref{sec:single_mode_model}. In the following, we choose $f(\phi)=0$ and $s=+1$ for the simplified potential
\begin{align}\label{eq:simple_potential}
    U_\mathrm{simple}(\phi)&=-E_{J\Sigma}\cos\theta\sqrt{1-\lambda\sin^2\frac{\phi}{2}},
\end{align}
which explicitly decouples the $(\phi,\theta)$-boundary condition in Eq.~\eqref{eq:coupled_boundary_condition}, making $\phi$ and $\theta$ $2\pi$-periodic independently. This can be exemplified for $k=d=0.2$ ($\lambda=0.96$) by comparing panels (a) and (b) of Fig.~\ref{fig:fig2_2d_potentials}. In panel (b), we plot $U_\mathrm{simple}$ where the line along the minima (solid green) is independent of $\phi$. We can therefore understand the simplified potential in Eq.~\eqref{eq:simple_potential} as shifting the minima of the fast mode to zero (mod $2\pi$) for all $\phi$. Furthermore, the simplified potential also directly admits the solution to the classical potential in Eq.~\eqref{eq:U_classical} by freezing $\theta$ to its minimum value corresponding to $\cos(\theta)=1$.

We now turn to the coupling coming from $f(\phi)$ in Eq.~\eqref{eq:U_prime} to gain intuition about which contributions are neglected in Eq.~\eqref{eq:simple_potential} by assuming that the fast mode is instantaneously centered around $f(\phi)$. For small $\phi$, $f(\phi)\approx (k-d)\phi/2$ which gives rise to a linear coupling between $\phi$ and $\theta$ that depends on the \emph{relative} asymmetry $k-d$ between capacitors and junctions. This is seen by expanding $\cos(\theta-f(\phi))$ from Eq.~\eqref{eq:U_prime} around the minimum at $\phi=\theta=0$
\begin{equation}\label{eq:phi_theta_coupling}
    \cos(\theta-f(\phi))\approx1-\frac{\theta^2}{2}+(k-d)\frac{\phi\theta}{2}-(k-d)^2\frac{\phi^2}{8}.
\end{equation}
The terms proportional to $\phi$ are neglected when the fast mode is assumed to instantaneously localize at the minimum set by $f^\phi$. Hence, we expect the BO approximation to be accurate for $k-d\approx0$ as also observed experimentally~\cite{shagalov2025double_junction_higher_harmonics, feldstein_bofill2026cooper_pair_parity}.

In Fig.~\ref{fig:fig2_2d_potentials}(c), we illustrate the coupling between the qubit and the internal mode by plotting $f(\phi)$ in Eq.~\eqref{eq:f_phi} for a fixed capacitance asymmetry $k=0.2$ and a varying junction asymmetry $d$. For $k=d$ (green line) $f(\phi)\approx0$ in an extended region around $\phi=0$, minimizing the coupling between the two modes (same parameters as panel (a) and (b)). Away from this point, small relative asymmetries $d=0$ and $d=0.5$ (dark and light green) give rise to a weak $\phi$-dependence resulting in a small coupling in Eq.~\eqref{eq:phi_theta_coupling}. For a strong relative asymmetry $d=-k=-0.2$ (red line) the two modes are strongly coupled as witnessed by the strong $\phi$-dependence in panel (d). In Appendix~\ref{app:no_coupling_BO}, we provide further numerical analysis of the errors arising from neglecting the coupling due to $f(\phi)$.

In Fig.~\ref{fig:fig2_2d_potentials}(d), we finally plot the potential energy along the minima (solid green) and maxima (dashed green) lines for different junction asymmetries to connect these to $U_\mathrm{semi}$ (same parameters as in panel (c)). The potential along the minima line is the classically obtained energy-phase relation $U_\mathrm{classical}(\phi)$ while the line along the maxima is the unstable classical equilibrium of the $\theta$-mode. Here, we directly observe the striking analogy to semiconductor-based junctions; the line along the minima corresponds to $U_\mathrm{semi}$ and the (unstable) line of maximal potential energy corresponds to the particle-hole symmetric branch $-U_\mathrm{semi}$.

\subsection{Effective single-mode model}\label{sec:single_mode_model}

We are now in a position to perform the BO approximation of the double-junction circuit. We separate the simplified potential in Eq.~\eqref{eq:simple_potential} into slow and fast potentials to uncover the slow and fast Hamiltonians
\begin{align}
    H_\mathrm{slow} &=4E_C(n-n_g)^2- E_{J\Sigma}\sqrt{1-\lambda\sin^2\frac{\phi}{2}},  \label{eq:H_slow_qubit}\\ 
    H_\mathrm{fast}^\phi &=4E_C^\mathrm{int}(N-N_g)^2 + E_{J,\mathrm{eff}}^\phi\left( 1-\cos\theta \right), \label{eq:H_fast_int}
\end{align}
where
\begin{equation}\label{eq:EJ_eff_internal}
    E_{J,\mathrm{eff}}^\phi = E_{J\Sigma}\sqrt{1-\lambda\sin^2\frac{\phi}{2}},
\end{equation}
is the effective Josephson energy of the internal mode which depends parametrically on $\phi$ and gives $E_{J,\mathrm{eff}}^{\phi=0}=E_{J\Sigma}$ consistent with the internal mode experiencing the junctions in parallel. Similarly, an effective Josephson energy of the qubit mode can be found from the curvature of the slow Hamiltonian, $E_{J,\mathrm{eff}}^\mathrm{slow}=\lambda E_{J\Sigma}/4=E_{J1}E_{J2}/(E_{J1}+E_{J2})$, consistent with the qubit mode experiencing the junctions in series. 

In the regime $E_{J\Sigma}\gg E_C^\mathrm{int}$, the internal mode is localized near the origin and we may approximate $1-\cos\theta\approx\theta^2/2$ and neglect the internal offset charge $N_g$. In the harmonic approximation, we obtain the ground state energy of the internal mode, giving the BO correction potential
\begin{equation}\label{eq:U_corr}
    U_\mathrm{corr}(\phi)= E_{J\Sigma}\sqrt{\frac{2E_C^\mathrm{int}}{E_{J\Sigma}}\sqrt{1-\lambda\sin^2\frac{\phi}{2}}},
\end{equation}
and the total BO Hamiltonian becomes
\begin{align}\label{eq:H_BO_final}
    H_\mathrm{BO} &= 4E_C \left(n-n_g\right)^2 +U_\mathrm{BO}(\phi),
\end{align}
where
\begin{equation}\label{eq:U_BO}
        U_\mathrm{BO}(\phi)=U_\mathrm{classical}(\phi)+U_\mathrm{corr}(\phi). 
\end{equation}
This effective single-mode Hamiltonian is a central result of our work and includes renormalization due to the internal mode through $U_\mathrm{corr}$ and junction capacitance asymmetry through of $E_C$ and $E_C^\mathrm{int}$, thus generalizing the results of Refs.~\cite{bozkurt2023fourier_engineering,rymarz2023singular_quantization}. Although this result is obtained in the regime $E_{J\Sigma}\gg E_C^\mathrm{int}$, we note that the ground state energy of the fast Hamiltonian in Eq.~\eqref{eq:H_fast_int} could also be solved numerically or using Mathieu functions, as is known from the transmon~\cite{koch2007transmon}. However, in the non-harmonic regime $E_{J\Sigma}/ E_C^\mathrm{int}<30$, the internal mode disperses with the offset charge on the middle island $N_g$. As we further elaborate in Sec.~\ref{sec:offset_charges}, this charge dispersion then propagates to the correction potential $U_\mathrm{corr}$, resulting in the qubit dispersing significantly with $N_g$. To avoid this internal mode-mediated dephasing channel, we focus on the parameter regime $E_{J\Sigma}/ E_C^\mathrm{int}\gtrsim30$, which also justifies neglecting the offset charge $N_g$ in $U_\mathrm{corr}$. In Appendix~\ref{Sec:num_BO}, we find that there is no difference in the accuracy of the BO approximation obtained from the harmonic approximation compared to a numerical solution when $E_{J\Sigma}/ E_C^\mathrm{int}\gtrsim30$.

In the next section, we analyze the numerical accuracy of the effective single-mode model, its harmonic content, and its sensitivity to offset charge.

\section{Results}\label{sec:results}

\subsection{Accuracy of the Born-Oppenheimer approximation}\label{sec:BO_accuracy}

We now turn to a detailed analysis of the numerical accuracy of the BO single-mode model and compare its performance to the classical single-mode model. We evaluate the numerical accuracies of the single-mode models by using the full two-mode model in Eq.~\eqref{eq:potential_phi1_phi2} as a reference. For this purpose, we define the cumulative relative error $\Delta_j=\sum_{i=1}^j\delta_{i}$ of the lowest $j$ qubit levels. Here, $\delta_{i}=|E_{i}'-E_{i}|/E_{i}$ is the relative error of the energy difference of the $i$'th qubit level and the ground state as computed using the two-mode model $E_i$ and a single-mode model $E_i'$. In the following, we explore how well the classical and BO single-mode models estimate the low-energy qubit spectrum for varying circuit parameters using the cumulative error $\Delta_j$.

We vary the circuit parameters while ensuring experimentally relevant parameters that places the qubit and internal mode frequencies in the BO regime. To do so, we set the charging energy of the qubit $E_C/h=200$\,\si{\mega\hertz} and demand that the qubit is effectively in a transmon regime of $E_J/E_C=50$. The effective Josephson energy of the qubit mode can be computed from Eq.~\eqref{eq:U_prime} or  Eq.~\eqref{eq:simple_potential} which, together with the transmon requirement, yields the condition $\lambda E_{J\Sigma}/4 = 50E_C$, specifying $E_{J\Sigma}$ for a given $\lambda$. These conditions also determine the qubit frequency which then scale as $E_{01}^\mathrm{qubit}\sim1/\sqrt{\lambda}$. For our parameters, the qubit frequency is approximately $f_q=3.7\,\si{\giga\hertz}$ for $\lambda=1$. We then determine $E_C^\mathrm{int}$ by choosing a transmon-like regime for the internal mode $E_{J\Sigma}/E_C^\mathrm{int} = 32$. Thus, for any $\lambda$, the internal mode frequency scales as $E_{01}^\mathrm{int} \sim 1/\lambda$, meaning that $E_{01}^\mathrm{qubit} < E_{01}^\mathrm{int}$ for all $ \lambda \leq 1$ as required for the validity of the BO approximation.

\begin{figure}
    \centering
    \includegraphics[width=1\columnwidth]{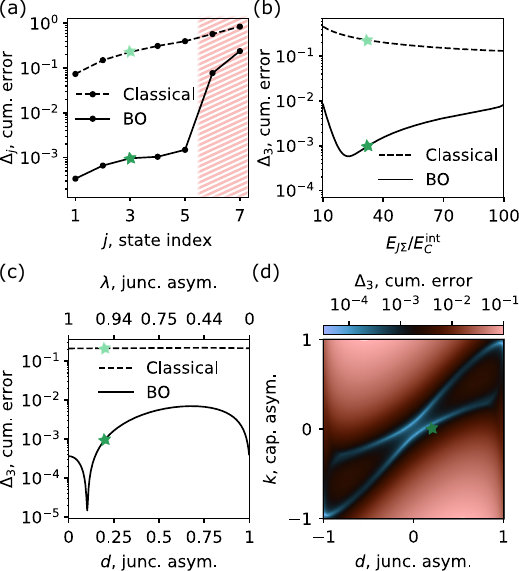}
    \caption{Comparison of the numerical accuracy of the classical (Eq.~\eqref{eq:U_classical}) and the BO (Eq.~\eqref{eq:U_BO}) model. (a) Cumulative error $\Delta_j$ relative to the full two-mode model (Eq.~\eqref{eq:potential_phi1_phi2}) of the lowest $j$ qubit levels, as defined in the main text (Sec.~\ref{sec:BO_accuracy}). The first excitation in the internal mode lies between the 5th an 6th qubit excitations and the breakdown of the BO model is observed for the states above it (red shaded region). (b-d) Cumulative error of the 3 lowest qubit levels ($j=3$) as a function of (b) the junction $\lambda$-parameter, (c) the $E_{J\Sigma}/E_C^\mathrm{int}$ ratio, and (d) the capacitance asymmetry $k$. The cumulative error of the BO model remains below $10^{-2}$ for most parameter choices and is roughly one to two orders of magnitudes less than the error of the classical model. Stars indicate points of identical configurations across all panels; $j=3$, $\lambda=0.96$, $E_{J\Sigma}/E_{C}^\mathrm{int}=32$, and $k=0.2$. Furthermore, $E_C/h=200$\,\si{\mega\hertz} and $\lambda E_{J\Sigma}/4 = 50E_C$ are fixed for all configurations.
    }
    \label{fig:BO_accuracy}
\end{figure}

In Fig.~\ref{fig:BO_accuracy}(a), we first plot the cumulative error for the classical and BO models as we increase the number of states that are being taken into account by $\Delta_j$. For the BO model (solid green line), the lowest 5 excitations are well described as the cumulative relative error is on the $10^{-3}$ level using the same fixed parameters as in Fig.~\ref{fig:fig2_2d_potentials} (indicated with stars in all panels). Compared to the classical model (dashed line), the BO model is about two orders of magnitude better in estimating the low-energy qubit spectrum. However, for the higher energy levels (6th and 7th) the cumulative error significantly increases for the BO model. This is because the BO approximation breaks down for levels above the first excited level of the internal mode (red shaded region), which here lies between the 5th and 6th qubit levels. In Appendix~\ref{Sec:num_BO}, we resolve the individual relative error for each level and find that they are very similar.

In Fig.~\ref{fig:BO_accuracy}(b-c), we plot the cumulative error for the lowest three excitations in the qubit for varying internal charging energy $E_C^\mathrm{int}$ and junction $\lambda$-parameter. Across both panels, the numerical accuracy of the BO model is on the order of $10^{-3}$ to $10^{-2}$ which is about one to two orders of magnitude more accurate than the classical model and corresponds to errors of a few to tens of $\si{\mega\hertz}$. In panel (b), we vary the charging energy of the internal mode, which displays an interesting minimum point of the cumulative error for the BO model. For small $E_{J\Sigma}/E_C^\mathrm{int}$, the harmonic approximation for the internal mode breaks down, leading to increased error. On the other hand, a large $E_{J\Sigma}/E_C^\mathrm{int}$ ratio for fixed $E_{J\Sigma}$, decreases the frequency of the internal mode, making its non-perturbative contribution more relevant despite the BO correction being less important, see Eq.~\eqref{eq:U_corr} and further discussion in Appendix~\ref{Sec:num_BO}. 
In panel (c), we fix the capacitance asymmetry $k=0.2$ and find that the BO model is exceptionally accurate near $d=0.2$, corresponding to no \emph{relative} asymmetry $k-d=0$. We interpret this as the qubit and internal mode only coupling to second order through the potential at this point since $f(\phi)\approx (k-d)\phi/2=0$, see Eqs.~\eqref{eq:U_prime}-\eqref{eq:f_phi}. 

In Fig.~\ref{fig:BO_accuracy}(d), we plot the cumulative error while simultaneously varying the junction asymmetry $d$ and the capacitance asymmetry $k$. Here, the BO model captures the low-energy spectrum exceptionally well around the diagonal $d=k$, consistent with the cancelation of the coupling due to $f(\phi)$. We additionally observe a vanishing error close to $k=0$, which we interpret as arising from cancelation between contributions in $\cos(\theta-f(\phi))$, see Appendix~\ref{app:no_coupling_BO} for further numerical analysis showing that this is a general feature. In conclusion, we find that the BO approximation accurately predicts the low-energy spectrum for all qubit states below the first excited internal state, especially near $k=0$ where the junction asymmetry can be tuned across a wide range. Accurate analytical results may be found in regions of large relative asymmetry through complementary analytical approaches such as Bogoliubov transformations~\cite{blais2021cqed_review}.

\subsection{Higher Josephson harmonics}
\begin{figure}
    \centering
    \includegraphics[width=1\linewidth]{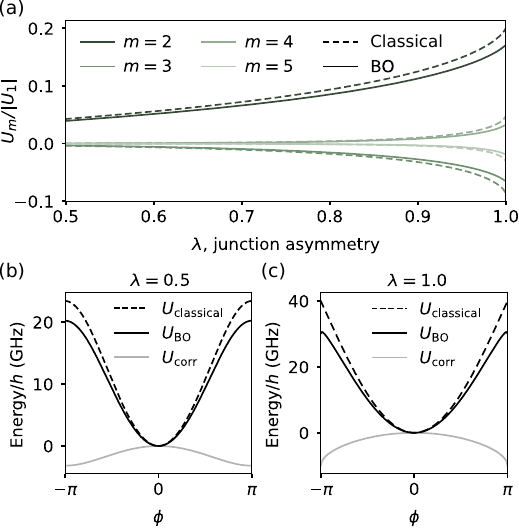}
    \caption{Harmonics of the single-mode classical and BO potentials. (a) The Josephson harmonic coefficients of the single-mode potential energy in the classical (dashed lines) and BO (solid lines) approximations, showing slightly reduced harmonics for the BO model. Coefficients are scaled to the fundamental frequency coefficient. (b,c) Comparison between the classical (black, dashed), BO (black, solid) and correction (grey, solid) potentials for (b) $\lambda=0.5$ and (c) $\lambda=1$. The correction potential becomes increasingly significant for higher energies, consistent with giving a larger correction to higher energy levels. Parameters are $E_{C}/h = 200$\,\si{\mega\hertz}, $\lambda E_{J\Sigma} / 4 = 50 E_C$, $E_{J\Sigma}/E_C^\mathrm{int} = 32$, and $k=0$ so $E_C^\mathrm{int}/h = 1.25$\,\si{\giga\hertz} for $\lambda=1$ and $E_C^\mathrm{int}/h = 2.5$\,\si{\giga\hertz} for $\lambda=0.5$.}
    \label{fig:harmonics}
\end{figure}

With the effective single-mode model developed in the previous section, we can examine the harmonic content of the potential in the double-junction circuit. The Josephson harmonic coefficients $U_m$ entering Eq.~\eqref{eq:Fourier_series} of a potential $U(\phi)$ can be found through
\begin{equation} \label{eq:harmonic_coefficients}
    U_m = \frac{1}{\pi}\int_{-\pi}^\pi  U(\phi)\cos(m\phi)\, d\phi.
\end{equation}
In Fig.~\ref{fig:harmonics}(a), we show the higher harmonic coefficients $U_m/|U_1|$ of the classical model Eq.~\eqref{eq:U_classical} (dashed lines) and the BO potential Eq.~\eqref{eq:U_BO} (solid lines) as a function of the asymmetry parameter $\lambda$. The presence of the fast internal mode, treated by the BO approximation, results in an overall reduction of the higher harmonic content. The second harmonic coefficient $U_2/|U_1|$ reaches a maximum of $0.17$ compared to the maximum classical value of $0.2$ at $\lambda=1$. 

In Fig.~\ref{fig:harmonics}(b-c), we further visualize the correction energy $U_\mathrm{corr} = U_\mathrm{BO} - U_\mathrm{classical}$ along with the potential energy seen by the qubit mode $\phi$ within the two models for (b) $\lambda=0.5$ and (c) $\lambda=1$. The potential barrier energy increases significantly with $\lambda$ in both models, but $U_\mathrm{corr}$ acts to lower the barrier height in the BO model and alters the $\phi$-dependence and, consequently, the harmonic content as seen in panel (a).

In this section, we include the offset charge on the middle island $N_g$ in the BO analysis to evaluate its impact on the qubit mode. In particular, we study the charge dispersion of the qubit to ensure that it is not limited by the potential dephasing induced by fluctuations in $N_g$.

The impact of $N_g$ on the fast internal mode is described by a transmon Hamiltonian, see Eq.~\eqref{eq:H_fast_int}. As is known from the transmon, the eigenenergies $E_m$ disperse with the offset charge according to
\begin{equation}\label{eq:energy_dispersion}
    E_m\approx E_m(N_g=0)+\varepsilon_m\sin^2(\pi N_g),
\end{equation}
where $\varepsilon_m$ is the peak-to-peak value of the charge dispersion~\cite{koch2007transmon}, see Fig.~\ref{fig:fig_5_charge_dispersion}(a) for the simulated charge dispersion of low-energy states at $N_g$ for $E_{J\Sigma}/E_C^\mathrm{int} = 8$. Here, the ground state energy of the internal mode (brown) is computed from Eq.~\eqref{eq:H_fast_int} at $\phi=0$ and displays harmonic dispersion as described by Eq.~\eqref{eq:energy_dispersion}. We also observe the ground (blue) and excited (orange) qubit energies obtained from the full two-mode model in Eq.~\eqref{eq:potential_phi1_phi2} to disperse in a similar fashion despite not being directly coupled to the charge on the central island $N_g$. 

\subsection{Sensitivity to offset charge}\label{sec:offset_charges}
\begin{figure}
    \centering
    \includegraphics[width=1\columnwidth]{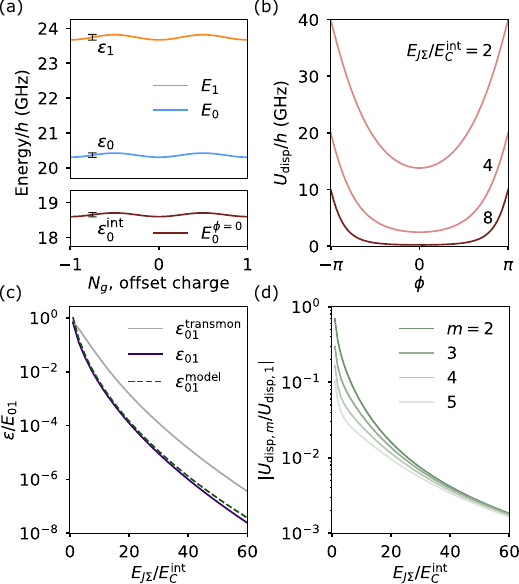}
    \caption{Numerically obtained charge dispersion of the qubit mediated by the internal mode. (a) Bottom: Charge dispersion of the internal mode ground state (brown) of the fast Hamiltonian (Eq. \eqref{eq:H_fast_int} at $\phi=0$) showing peak-to-peak value $\varepsilon_0^\mathrm{int}$. Top: Charge dispersion of the qubit ground (blue) and first excited (orange) energies of the full two-mode model (Eq.~\eqref{eq:potential_phi1_phi2}) showing inherited dependence on the offset charge $N_g$ from the internal mode. Parameters used are $E_{J\Sigma}/E_C^\mathrm{int} = 8$ to illustrate pronounced $N_g$-dependence. (b) The dispersive potential (Eq.~\eqref{eq:dispersion_potential}) shown for different $E_{J\Sigma}/E_C^\mathrm{int}$ ratios, displaying exponential suppression for increasing ratios. (c) Peak-to-peak charge dispersion versus $E_{J\Sigma}/E_C^\mathrm{int}$ of the qubit (purple, solid) computed from the full two-mode model (Eq.~\eqref{eq:potential_phi1_phi2}) compared to our first-order model (green, dashed) using the dispersive potential (Eq.~\eqref{eq:perturbation_theory}). To evaluate $\varepsilon_{01}^\mathrm{model}$ we use eigenstates of the BO Hamiltonian (Eq.~\eqref{eq:H_BO_final}). We also compare to the charge dispersion of a transmon (grey), observing orders-of-magnitude less charge dispersion of the qubit mode. (d) Fourier coefficients of the dispersion potential $U_\mathrm{disp}$, showing suppression below $10^{-2}$ for $E_{J\Sigma}/E_C^\mathrm{int}\gtrsim30$.  For all panels, $\lambda = 1$, $E_C/h = 200$\,\si{\mega\hertz}, $E_{J\Sigma}/h  = 40$\,\si{\giga\hertz}, $k=0$.}
    \label{fig:fig_5_charge_dispersion}
\end{figure}

To understand the qubit's dispersion with $N_g$, we note that, in the context of the BO approximation, the ground state energy $E_0$ corresponds to the correction potential $U_\mathrm{corr}$. The BO correction potential therefore acquires an offset charge dependence
\begin{equation}\label{eq:charge-dependent-correction}
    U_\mathrm{corr}(N_g)\approx U_\mathrm{corr}(N_g=0)+U_\mathrm{disp}\sin^2(\pi N_g),
\end{equation}
where 
$U_\mathrm{corr}(N_g=0)$ is given by Eq.~\eqref{eq:U_corr} and $U_\mathrm{disp}$ is the contribution due to the peak-to-peak value of the charge dispersion $\varepsilon_0$.
This ``dispersion'' potential contains two parts
\begin{equation}\label{eq:dispersion_potential}
    U_\mathrm{disp}(\phi) = \varepsilon_0^\mathrm{int}\, u_\mathrm{disp}(\phi),
\end{equation}
where the $\phi$-dependency is collected in
\begin{equation}
    u_\mathrm{disp}(\phi)=g(\phi)^{\frac{3}{4}} \exp\left[{-\sqrt{\frac{8E_{J\Sigma}}{E_C^\mathrm{int}}}\left(\sqrt{g(\phi)}-1\right)}\right],
\end{equation}
with $g(\phi)=\sqrt{1-\lambda\sin^2(\phi/2)}$ and where
\begin{equation}
    \varepsilon_0^\mathrm{int}\approx 2^5 \sqrt{\frac{2}{\pi}}E_C^\mathrm{int} \left(\frac{E_{J\Sigma}}{2E_C^\mathrm{int}}\right)^{3/4} e^{-\sqrt{8E_{J\Sigma}/E_C^\mathrm{int}}},
\end{equation}
is an exponentially suppressed prefactor corresponding to the ground state charge dispersion of a transmon with charging energy $E_C^\mathrm{int}$ and Josephson energy $E_{J\Sigma}$~\cite{koch2007transmon}. 

In Fig.~\ref{fig:fig_5_charge_dispersion}(b), we plot the dispersion potential in Eq.~\eqref{eq:energy_dispersion} for different values of $E_{J\Sigma}/E_C^\mathrm{int}$. We directly observe how the exponentially small prefactor $\varepsilon_0^\mathrm{int}$ suppresses the dispersion potential for increasing $E_{J\Sigma}/E_C^\mathrm{int}$. This is especially pronounced around $\phi=0$ which is the relevant region for low-energy qubit states.

To quantitatively assess the impact of the dispersion potential on the qubit energy, we treat $U_\mathrm{disp}$ as a perturbation to the BO Hamiltonian Eq.~\eqref{eq:H_BO_final}. First-order perturbation theory then leads to a correction to the qubit energies and consequently an expression for the charge dispersion of the qubit mode due to $N_g$
\begin{equation}\label{eq:perturbation_theory}
    \varepsilon_{01}^\mathrm{model} = \varepsilon_0^\mathrm{int}\left[ \braket{1|u_\mathrm{disp}(\phi)|1} - \braket{0|u_\mathrm{disp}(\phi)|0} \right],
\end{equation}
which is scaled by the exponentially suppressed $\varepsilon_0^\mathrm{int}$, inherited from the internal mode.

In Fig.~\ref{fig:fig_5_charge_dispersion}(c), we compare the peak-to-peak charge dispersion of the qubit energy obtained from the full two-mode model in Eq.~\eqref{eq:potential_phi1_phi2} (purple, solid) to a numerical solution of Eq.~\eqref{eq:perturbation_theory} (green, dashed) using wavefunctions obtained from the BO Hamiltonian Eq.~\eqref{eq:H_BO_final}. We find excellent agreement between the full solution and our first-order perturbative model, showing that the BO approximation can also be used to account for dephasing effects mediated by the internal mode. We also compare these results to the charge dispersion of a standard transmon (gray). We find that the sensitivity to the internal offset charge $N_g$ is suppressed to low-levels comparable to the low sensitivity of transmons for $E_{J\Sigma}/E_C^\mathrm{int}\gtrsim30$. Given the qualitative similarities between the effective model of the double-junction circuit in Eq.~\eqref{eq:H_BO_final} and the transmon, we believe that their coherence properties will otherwise be largely the same; however, a complete coherence analysis is beyond the scope of this work.

In Fig.~\ref{fig:fig_5_charge_dispersion}(d), we finally plot the lowest harmonics of the dispersion potential $U_\mathrm{disp}$ as a function of $E_{J\Sigma}/E_C^\mathrm{int}$. We find that they are sufficiently suppressed ($<10^{-2}$) for $E_{J\Sigma}/E_C^\mathrm{int}\gtrsim30$, meaning that the higher lying states and harmonics analysis is not impacted by the dispersion potential. In conclusion, our analysis shows that the offset charge on the middle island $N_g$ can be neglected for $E_{J\Sigma}/E_C^\mathrm{int}\gtrsim30$.

\section{Conclusions} \label{sec:conclusion}

We present an effective single-mode model to accurately describe the low-energy qubit spectrum of a double-junction element shunted by a large capacitor. Using a BO approximation, we obtain a correction term to the effective potential energy coming from zero-point fluctuations of the double-junction element's internal degree of freedom. We analyze the numerical accuracy of the BO model for experimentally relevant parameter regimes and find that the inclusion of the correction term decreases the error in the numerically obtained low-energy qubit spectrum by two orders of magnitude, resulting in low errors at the few to tens of $\si{\mega\hertz}$ level. Especially for symmetric junction capacitances, we find that the BO model accurately predicts the energy of all states below the first excited internal state while allowing for a large tunability in the junction asymmetry. Supporting these results are two parallel experimental works on double-junction qubits~\cite{shagalov2025double_junction_higher_harmonics, feldstein_bofill2026cooper_pair_parity}. In these works, the BO single-mode model obtained here is required to accurately fit the respective qubit spectra. In Ref.~\cite{shagalov2025double_junction_higher_harmonics} the relative junction energy of an all-SIS double-junction transmon is tuned in-situ via flux control to experimentally probe in which parameter regimes the effective single-mode model describes the observed data. In Ref.~\cite{feldstein_bofill2026cooper_pair_parity} the double-junction element is placed in parallel to a semiconductor-based junction such that interference between the two arms leads to controllable higher Josephson harmonics, illustrating how the double-junction element can be used as an effective single component. 

This work further extends the analysis to the offset charge dispersion of the internal degree of freedom. We find a transmon regime ($E_{J\Sigma}/E_C^\mathrm{int}\gtrsim30$) for the internal mode, which exponentially suppresses the sensitivity of the qubit mode to internal offset charges. These results enable us to analyze the harmonic content of the double-junction element, and we find that the correction term only slightly decreases the overall content of higher Josephson harmonics. Finally, our analysis reveals that a family of potential energies gives rise to the same BO correction term. This enables a simplified approach when performing the BO approximation; first, simplify the multi-mode potential within the family of interest, and second, perform the conventional steps in the BO approximation. In summary, our analysis and results show that the double-junction element can be readily used as a resource for higher Josephson harmonics in multi-mode circuits and further validate the BO approximation as a powerful tool for studying multi-mode superconducting circuits.

\begin{acknowledgments}
We gratefully acknowledge useful discussions with Clinton A.~Potts, Valla Fatemi, András Gyenis, Nataliia Zhurbina, Christian Kraglund Andersen, Adrián Parra-Rodriguez, Benjamin Leviatan, Alexandre Blais, Cheng-Li Chen, and Kohei Matsuura.

This research was supported by the Novo Nordisk Foundation (grant no. NNF22SA0081175), the NNF Quantum Computing Programme (NQCP), Villum Foundation through a Villum Young Investigator grant (grant no. 37467), the Innovation Fund Denmark (grant no. 2081-00013B, DanQ), the U.S. Army Research Office (grant no. W911NF-22-1-0042, NHyDTech), by the European Union through an ERC Starting Grant, (grant no. 101077479, NovADePro), and by the Carlsberg Foundation (grant no. CF21-0343). 
Any opinions, findings, conclusions or recommendations expressed in this material are those of the author(s) and do not necessarily reflect the views of Army Research Office or the US Government. 
Views and opinions expressed are those of the author(s) only and do not necessarily reflect those of the European Union or the European Research Council. Neither the European Union nor the granting authority can be held responsible for them. 
Finally, we gratefully acknowledge Lena Jacobsen and Helle Grunnet for program management support.
\end{acknowledgments}

\appendix
\setcounter{figure}{0}
\renewcommand{\thefigure}{A\arabic{figure}}

\section{Errors in the Born-Oppenheimer approximation}\label{app:BO_errors}
In this appendix, we perform further numerical analysis to understand the small errors found in Sec.~\ref{sec:BO_accuracy} coming from the BO approximation.

\begin{figure}
    \centering
    \includegraphics[width=\linewidth]{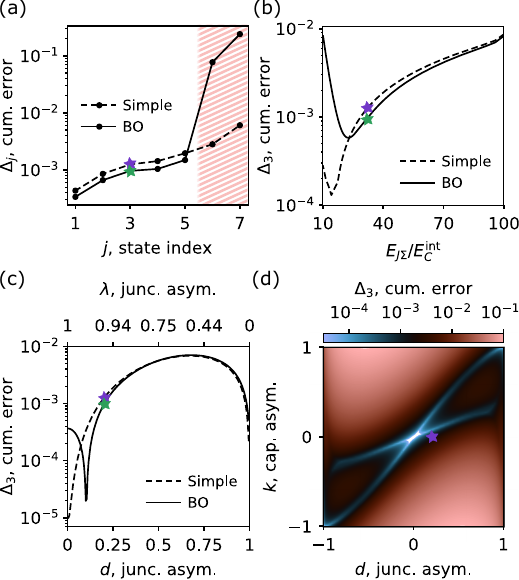}
    \caption{Comparison of the accuracy of the simplified two-mode model (Eq.~\eqref{eq:simple_potential}) and the BO model (Eq.~\eqref{eq:H_BO_final}). Across all panels, we observe very similar errors of the simplified (dashed line) and BO (solid line) models in the regime where the BO approximation is expected to work (further details in main text). Parameters used are identical to Fig.~\ref{fig:BO_accuracy}.}
    \label{fig:2D_comparisons}
\end{figure}

\subsection{BO errors and the internal mode}\label{app:no_coupling_BO}

\begin{figure}
    \centering
    \includegraphics[width=\linewidth]{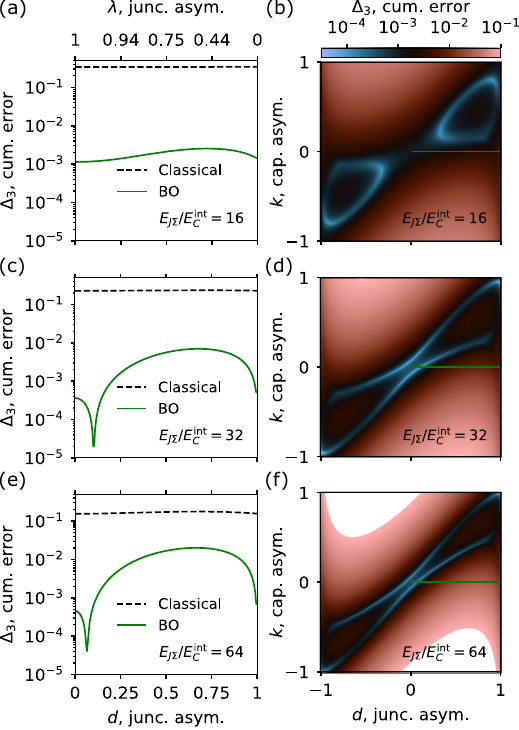}
    \caption{Cumulative error in the BO model (Eq.~\eqref{eq:H_BO_final}) for different values of $E_{J\Sigma}/E_C^\mathrm{int}$. (a,c,e) Cumulative error of the classical (black) and BO (green) models for $E_{J\Sigma}/E_C^\mathrm{int}=16,\, 32,\, 64$. (b,d,f) Cumulative error of the BO model as a function of junction and capacitance asymmetry (green line indicating line cut in left panels). We observe slightly increasing errors for increasing $E_{J\Sigma}/E_C^\mathrm{int}$ (further details in main text). Parameters used are identical to Fig.~\ref{fig:BO_accuracy}.}
    \label{fig:BO_for_different_EJ_EC}
\end{figure}

In this section, we investigate the origin of the remaining small errors in the BO approximation in a broad parameter regime of the internal mode. We show how these errors can be understood as coming from the approximation that the fast (internal) mode instantaneously localizes at the minimum set by the slow (qubit) mode. To see this, we compare the error of the simplified two-mode model in Eq.~\eqref{eq:simple_potential}  to the BO error as studied in Sec.~\ref{sec:BO_accuracy}. The simplified two-mode model removes the $\phi$-dependence of the internal mode's minimum by setting $f(\phi)=0$ without integrating the internal mode out. In this way, we can compare the errors that arise from assuming the instant localization of the internal mode and the other approximations made when integrating out the internal mode.

In Fig.~\ref{fig:2D_comparisons}, we numerically evaluate the cumulative error of the simplified two mode model (Eq.~\eqref{eq:simple_potential}) with respect to the full model (Eq.~\eqref{eq:potential_phi1_phi2}) and compare it to the BO model (Eq.~\eqref{eq:H_BO_final}) as in Sec.~\ref{sec:BO_accuracy}. Across all panels, we observe that the error in the two models match very well except where we expect the BO approximation to fail (stars indicate identical parameters); in panel (a), we find that the simplified model (solid line) accurately predicts the high-energy qubit energies above the first excitation in the internal mode, and in panel (b), we observe that the simplified model also describes the qubit spectrum at low $E_{J\Sigma}/E_C^\mathrm{int}$ since no harmonic approximation is made. In panels (c,d), the error in the simplified two-mode model is strikingly similar to the BO model. This comparison highlights that the errors in the BO approximation are mainly due to the assumption that the internal mode instantaneously localizes near the minimum set by $f(\phi)$ (Eq.~\eqref{eq:f_phi}) and not by integrating the internal mode out.

With this understanding, we now turn to studying how the BO error depends on varying the $E_{J\Sigma}/E_C^\mathrm{int}$ ratio of the internal mode. In Fig.~\ref{fig:BO_for_different_EJ_EC}, we plot the error of the classical (black) and BO (green) models as in Sec.~\ref{sec:BO_accuracy} for $E_{J\Sigma}/E_C^\mathrm{int}=16,\, 32,\, 64$ as the junction and capacitance asymmetries are varied. In panels (a,c,e), we show a line cut at $k=0$ and observe that errors in the BO model are on the $10^{-3}-10^{-2}$ level compared to $>10^{-1}$ for the classical model. In panels (b,d,f), we map out the BO error for all possible values of junction and capacitance asymmetry (green line indicates line cut in left panels). For increasing $E_{J\Sigma}/E_C^\mathrm{int}$, we observe the low-error regions in the lower left and upper right are shrinking. This is consistent with the interpretation that the BO error originates from the internal mode not localizing instantaneously at the minimum set by the qubit mode; increasing $E_{J\Sigma}/E_C^\mathrm{int}$ decreases the frequency of the internal mode (we fix the qubit frequency and $E_{J\Sigma}$ for a fixed value of $d$ and $k$), thus making it slower and increasing the errors due to the fast approximation. These results support our analysis of the origin of errors in the BO model and show that the BO model can predict the low-energy qubit spectrum for symmetric capacitances $k=0$ in a broad parameter regime.

\subsection{Numerical Born-Oppenheimer approximation}\label{Sec:num_BO}

In this appendix, we analyze the numerical solution of the fast Hamiltonian Eq.~\eqref{eq:H_fast_int} and compare it to the analytically obtained solution in Eq.~\eqref{eq:U_corr}. The numerical solution refers to solving for the ground state energy of the fast mode in Eq.~\eqref{eq:H_fast_int} for an array of values $\phi$. This ground state energy then constitutes the correction potential in Eqs.~\eqref{eq:U_corr} and~\eqref{eq:H_BO_final}. Using this numerically obtained correction potential, the single-mode qubit states and energies are then computed numerically as for the analytically obtained correction potential.

\begin{figure}
    \centering
    \includegraphics[width=\linewidth]{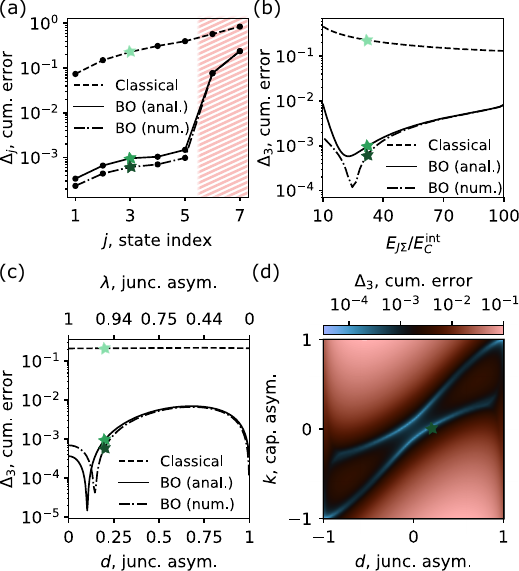}
    \caption{Comparison of the accuracy of the single-mode BO Hamiltonian (Eq.~\eqref{eq:H_BO_final}) when the fast Hamiltonian (Eq.~\eqref{eq:H_fast_int}) is solved analytically (Eq.~\eqref{eq:U_corr}) or numerically. Across all panels we observe good agreement between the analytical (solid line) and numerical (dash-dotted line) BO solution (further details in main text). Parameters used are identical to Fig.~\ref{fig:BO_accuracy}.}
    \label{fig:comparison_anal_BO}
\end{figure}

In Fig.~\ref{fig:comparison_anal_BO}, we compare the accuracy of the numerical BO model with the results of Fig.~\ref{fig:BO_accuracy}. Across all panels, we observe that the accuracy of the numerical (dash-dotted line) and analytical (solid line) BO models is very similar (stars indicate identical parameters). Only in panel (c) for values of $E_{J\Sigma}/E_C^\mathrm{int}$ below 30, does the numerical model slightly outperform the analytical model. This is a consequence of the harmonic approximation used in the analytical model failing for lower values of $E_{J\Sigma}/E_C^\mathrm{int}$. The numerical solution performs better as it directly solves for the ground state energy without making any additional assumptions. However, as argued in Sec.~\ref{sec:offset_charges}, the practical parameter regime of interest lies at $E_{J\Sigma}/E_C^\mathrm{int}\gtrsim30$ and for this reason we focus on the simpler analytical BO model in this work.

\begin{figure}
    \centering
    \includegraphics[width=\linewidth]{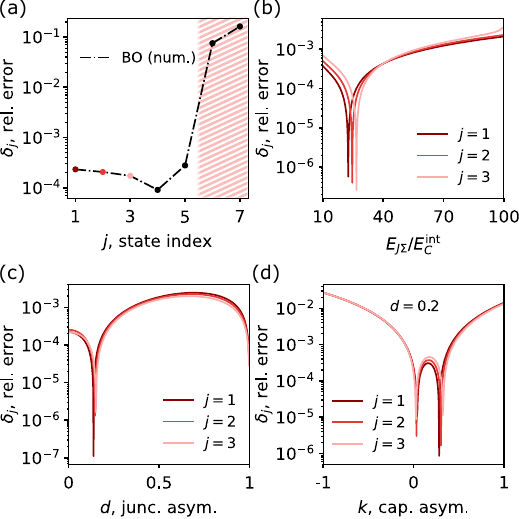}
    \caption{The individual relative errors $\delta_j$ of the lowest levels in the numerical BO model. (a) Relative error for the seven lowest qubit excitations. The energy of excitations 1-5 are well-described by the BO approximation as they are below the first excitation in the internal mode. (b-d) Individual relative error of the first three qubit excitations as a function of $E_{J\Sigma}/E_C^\mathrm{int}$, junction asymmetry $d$ and capacitance asymmetry $k$. The error for each of the states is very similar in the wide parameter range explored here. Parameters used are identical to Fig.~\ref{fig:BO_accuracy}.}
    \label{fig:separated_errors}
\end{figure}

In Fig.~\ref{fig:separated_errors}, we resolve the cumulative errors of the numerical solution in Fig.~\ref{fig:comparison_anal_BO} and plot the individual relative error $\delta_j$ as defined in Sec.~\ref{sec:BO_accuracy}. In panels (a,b,c), we plot the relative error for varying state index, $E_{J\Sigma}/E_C^\mathrm{int}$ ratio, and junction asymmetry as in Fig.~\ref{fig:BO_accuracy}. In panel (d), we plot the relative error for the three lowest states as a function of the capacitance asymmetry for $d=0.2$ and note the striking low-error point at $k=d$. Across all panels, we observe very similar relative errors for the lowest three excitation energies, showing that the entire low-energy spectrum below the first excitation of the internal mode can be captured by the BO model.

\clearpage
\bibliography{references}

\end{document}